\documentclass[prb,twocolumn,showpacs,showkeys,amsmath]{revtex4}
\usepackage{graphicx}
\usepackage{amssymb}
\usepackage{bm}
\usepackage{kotex}
\begin{document}
\title{Finite entanglement properties in the matrix product states of the one-dimensional Hubbard model}
\author{Min-Chul Cha}
\affiliation{Department of  Photonics and Nanoelectronics, Hanyang University, Ansan, Gyeonggi-do 15588, Korea}
\begin{abstract}
We study the effects due to limited entanglement in the one-dimensional Hubbard model by representing the ground states in the form of the matrix product states.
Finite-entanglement scaling behavior over a wide range is observed at half-filling.
The critical exponents characterizing the length scale in terms of the size of matrices used are obtained,
confirming the theoretical prediction that the values of the exponents are solely determined by the central charge.
The entanglement spectrum shows that a global double degeneracy occurs in the ground states with a charge gap. 
We also find that the Mott transition, tuned by changing the chemical potential, always occurs through a first-order transition and the metallic phase has a few conducting states, including the states with the mean-field nature close to the critical point,
as expected in variational matrix product states with a finite amount of entanglement.
\end{abstract}

\maketitle

\section{Introduction} 
Entanglement is quantum correlations characterizing the inseparability between blocks of quantum states.
Because of this feature, entanglement is closely related with the ground state properties of strongly correlated systems
\cite{Amico08,Calabrese09,Eisert10,Laflorencie16}
which often undergo changes through quantum phase transitions.
It has been realized that the changes can be captured by the amount of entanglement in the wave functions of the ground states.\cite{Fazio02,Osborne02}
This idea develops to the scaling of entanglement entropy (EE) in critical systems with a relevant length scale,
\cite{Vidal03,Korepin04,Calabrese04,Holzhey94}
such as block size or correlation length.
In the systems with limited entanglement, the finite-entanglement scaling behavior\cite{Pollmann09}
can be obtained by representing the amount of entanglement in terms of a characteristic length scale.
The finite-entanglement scaling behavior at quantum phase transitions have been found
in various one-dimensional systems.\cite{Pollmann09,Tagliacozzo08,Pino12,Pirvu12,Wang15}
Recently it has been discussed that more information about entanglement of a quantum state can be obtained from the entanglement
spectrum (ES) distribution,\cite{Li08,Pollmann10s,Calabrese08,Alba18,Peschel09}
from which the finite-entanglement scaling theory can be derived.

The limitation in the amount of entanglement also affects the nature of transition near a quantum criticality.
The finite size of the critical systems smooths out the singularity by confining the correlation length,
which leads the size to be treated as a scaling variable that behaves with a correct critical exponent asymptotically.
Whether a finite amount of the entanglement, which often characterized by a finite length scale, brings an effect similar to
a finite size of the system in the vicinity of criticality is an interesting question, which is less studied.
Obviously, in the limit of diminishing entanglement, a mean-field-like behavior will govern the transition\cite{Liu10}.
It is a non-trivial question how the behavior of this transition changes, which is often characterized by its corresponding exponents,
as the amount of the entanglement increases.

In this work, we study the finite-entanglement properties of the Mott phase
and their effects on the metal-insulator quantum phase transition in the one-dimensional Hubbard model\cite{Korepin05Book}
by representing the ground state wavefuncion in matrix product states (MPS) form
\cite{Orus14, Schollwock11,PGarcia07,McCulloch07,Pirvu11,Pippan10,Vidal03b}.
The MPS approach provides a useful platform for investigating how a limited amount of entropy,
characterized by the matrix size  $\chi$, shows scaling behavior and affects the nature of the transition.
Based on the assumption that a finite amount of entanglement induces a length scale\cite{Pollmann09,Tagliacozzo08}
$\xi_\chi\sim\chi^\kappa$, 
where $\kappa$ is an exponent whose value is solely determined by the central charge $c$,
we investigate the scaling behavior over this length scale in the critical region. 
In addition, with a finite $\chi$, we find that the transition from the metal to the insulator occurs via a first-order transition.
We also find multiple variational MPS solutions in some regions of the metallic phase, 
including mean-field like solutions near the critical point,
and first-order transitions between them.

This paper is organized as follows.
In Section~II, we introduce the MPS method for the one-dimensional Hubbard model, which yields an accurate estimation
of the ground state energy in the process of the time-evolving block decimation (TEBD).\cite{Vidal04,Vidal07} 
Section~III, we discuss the finite-entanglement scaling and the properties of
the ES of the Mott phase. The MPS calculation for various $\chi$ is compared with the prediction of the conformal field theory (CFT).
In Section~IV, we investigate how the finite entanglement in the MPS changes the nature of quantum phase transitions. 
Finally, the conclusions are summarized in Sec.~V.

\section{MPS representations}
The Hubbard model on a one-dimensional infinite chain is given by the Hamiltonian
\begin{eqnarray}
H=&U&\sum_i(n_{i\uparrow}-\frac{1}{2})
(n_{i\downarrow}-\frac{1}{2})-\mu\sum_{i\sigma}n_{i\sigma}\cr
&-&t\sum_{i\sigma}(c_{i+1 \sigma}^\dagger c_{i\sigma}+c_{i\sigma}^\dagger c_{i\sigma}),
\label{eq:H}
\end{eqnarray}
where $i$ is the index for sites, $\sigma=\uparrow, \downarrow$  is the spin coordinates,
$U$ is the strength of the on-site interaction, $t$ is the hopping amplitude,  and $\mu$ is the chemical potential. $n_{i\sigma}$ represents the number operator and $c_{i\sigma}^\dagger (c_{i\sigma})$ denotes the creation (annihilation) operator at the $i$-th site for spin $\sigma$.
We take the energy unit $t=1$ and investigate the metal-insulator transition of the system by tuning
$\mu$ for various $U$.

Taking into account of its translational invariance,
a variational ground state of the system is constructed in the canonical form of the MPS wavefunction\cite{Vidal07}
\begin{eqnarray}
|\Psi^0\rangle =\sum_{\{s_i, a_i\}}
&&A^{[s_1]}_{a_L a_1}\Lambda_{a_1} B^{[s_2]}_{a_1a_2}\Lambda_{a_2}A^{[s_3]}_{a_2a_3}
\cdots B^{[s_L]}_{a_{L-1}a_L}\Lambda_{a_L}\cr
&&\times|s_1,s_2,s_3,\cdots,s_L\rangle,
\label{eq:MPS}
\end{eqnarray}
where, with $L\to\infty$ for an infinite chain, $A$'s and $B$'s are the MPS matrices of size $\chi \times \chi$, $s_i$ are physical indices for the basis states ($|s_i\rangle =|0\rangle,|\uparrow\rangle,|\downarrow\rangle,|\uparrow\downarrow\rangle$) of the $i$-th site,
and  $a_i=1,\cdots,\chi$ are the bond indices.
Here, we take two sets of matrices $A$'s  and $B$'s, multiplied alternatively, for the convenience in using
the two-site TEBD algorithm in the process seeking the lowest energy state.
Because of the translational invariance, $A$'s and $B$'s are different up to a gauge choice
(i.e. $A^{[s]}=QB^{[s]}Q^{-1}$ with a unitary matrix $Q$).
It turns out that the same column vector $\Lambda$, whose elements $\Lambda_a$ are the Schmidt coefficients,
appears at each site.

The coefficients $\Lambda_a$ are real non-negative numbers, ordered such that $\Lambda_1\ge\Lambda_2\ge\cdots$.
We adopt a normalization scheme $\Lambda_1=1$.
We choose a finite $\chi$ by truncating the Hilbert space 
because $\Lambda_a$ decays rapidly as $a$ increases for weak entanglement.
Larger $\chi$ is preferred, in general, for more accurate calculations dealing with strong entanglement.
In the critical region, the limited size $\chi$ induces systematic errors reflecting the amount of
entanglement included in the ground state wavefunction.
This leads us to expect scaling behavior as a function of $\chi$.

In order to determine $A$'s, $B$'s, and $\Lambda$ that minimize the energy of the variational wavefunction in
Eq.~(\ref{eq:MPS}), we use the imaginary TEBD method: Starting with an arbitrary initial $|\Psi\rangle$, we expect
\begin{eqnarray}
\lim_{\tau\to\infty} e^{-\tau H}|\Psi\rangle\to|\Psi^0\rangle,
\end{eqnarray}
where $|\Psi^0\rangle$ is the ground state.
Because $e^{-\tau H}$ is an operator containing non-commuting terms, the Suzuki-Trotter decomposition is applied
by dividing time into small intervals of size $\Delta\tau$.
Then, we have $e^{-\tau H}=(e^{-\Delta\tau H})^N (\Delta\tau=\tau/N)$.
When $\Delta\tau\ll 1$, the Suzuki-Trotter decomposition leads to
\begin{eqnarray}
e^{-\Delta\tau H}=\prod_{i=1,3,\cdots}^{L-1}e^{-\Delta\tau h_{i,i+1}}\prod_{i=2,4,\cdots}^Le^{-\Delta\tau h_{i,i+1}},\cr
h_{i,i+1}=\frac{1}{2}(h^0_i+h^0_{i+1})-t\sum_{\sigma}(c_{i+1 \sigma}^\dagger c_{i\sigma}+c_{i\sigma}^\dagger c_{i+1\sigma}),\cr
h^0_i=U(n_{i\uparrow}-\frac{1}{2})(n_{i\downarrow}-\frac{1}{2})-\mu(n_{i\uparrow}+n_{i\downarrow}),
\end{eqnarray}
where we bipartite the system into two parts containing odd and even bonds.
A smaller $\Delta\tau$ increases the accuracy of the decomposition, but requires longer time in the TEBD calculations.
We choose as small as $\Delta\tau=0.002$ in our calculations.

In a uniform system, because of the translational symmetry, 
we expect
\begin{eqnarray}
e^{-\Delta\tau h_{i,i+1}}|\Psi^0\rangle=e^{-\Delta\tau\varepsilon_0}|\Psi^0\rangle,
\label{eq:one-operation}
\end{eqnarray}
where $\varepsilon_0$ is the ground-state energy per site.
To find $\varepsilon_0$ numerically in the process of the two-site TEBD, we define a matrix
\begin{eqnarray}
\Theta^{s_1s_2}_{ab}=\sum_{a_1,s^\prime_1,si^\prime_2}
\langle s_1s_2|e^{-\Delta\tau h_{12}}|s^\prime_1s^\prime_2\rangle
\Lambda_a A^{[s^\prime_1]}_{aa_1}\Lambda_{a_1}B^{[s^\prime_2]}_{a_1b}\Lambda_b,
\label{eq:Theta}
\end{eqnarray}
for example, for an odd site, and rewrite it in the form
\begin{eqnarray}
\Theta^{s_1s_2}_{ab}=\sum_{\gamma=1}^{4\chi}
{\tilde U}_{\alpha\gamma}\Sigma_\gamma {\tilde V}_{\beta\gamma}
\end{eqnarray}
by using a singular-value decomposition.
Here, ${\tilde U}$ and ${\tilde V}$ are $4\chi\times 4\chi$ unitary matrices with indices
$\alpha=\chi s_1+a$ and $\beta=\chi s_2+b$ ($s_1, s_2=0, 1, 2, 3$), and
$\Sigma$ is a column vector of size $4\chi$ with its elements arranged in decreasing order $\Sigma_1\ge\Sigma_2\ge\cdots$.
The time-evolved (updated) matrices and coefficients are then obtained via redefining
${\tilde A}^{[s_1]}_{aa_1}={\tilde U}_{\alpha a_1}/\Lambda_a,
{\tilde B}^{[s_2]}_{a_1b}={\tilde V}_{\beta a_1}/\Lambda_b$,
and ${\tilde\Lambda}_{a_1}=\Sigma_{a_1}/\Sigma_1$.
Repeating the same TEBD until $A$'s, $B$'s, and the $\Lambda$'s converge to the ground state
where these matrices and column vectors  remain unchanged.
Then, we can find in Eq.~(\ref{eq:one-operation}) that $\varepsilon_0$ is obtained from the norm of the updated wavefunction as
\cite{Cha15}
\begin{eqnarray}
\varepsilon_0=-\frac{1}{\Delta\tau}\ln{\Sigma_1}.
\label{eq:varepsilon_0}
\end{eqnarray}
We will find below that this method gives an accurate estimation of $\varepsilon_0$ for a given $\chi$.

\section{Entanglement Properties at $\mu=0$}
\subsection{scaling behavior}
The exact wavefunctions will be realized only in the limit $\chi \to \infty$.
A finite $\chi$ limits the amount of entanglement involved and introduces systematic errors.
These effects can be characterized by an effective quantum correlation length,
\begin{eqnarray}
\xi_\chi\sim \chi^\kappa,
\label{eq:xi_chi}
\end{eqnarray}
with an exponent $\kappa$.
This length scale roughly defines a range over which the entanglement between parts has to be counted.
It has been proposed\cite{Pollmann09}, based on the CFT,
that the value of $\kappa$ is determined only by the central charge $c$ in the form
\begin{eqnarray}
\kappa=\frac{6}{c(\sqrt{12/c}+1)}.
\label{eq:kappa}
\end{eqnarray}
The Hubbard model at half-filling ($\mu=0$) is an ideal place to check this form in a single model: 
for $U=0$, both the charge and the spin excitations are gapless, leading to $c=2$,
while for $U>0$ the charge fluctuations are gapped so that the central charge becomes $c=1$.
Eq.~(\ref{eq:kappa}) provides then $c\kappa/6=0.290\ (\kappa=0.870)$ for $c=2$
and $c\kappa/6=0.224\ (\kappa=1.344)$ for $c=1$.
Below we check these numbers from the scaling behavior of various quantities.

Because the quantum correlation length of the MPS ground state represented in Eq.~(\ref{eq:MPS}) is limited by $\xi_\chi$,
the energy per site (i.e. free energy density at zero temperature),
$\varepsilon_0(\chi)$, of the ground state with a finite $\chi$ obeys the following scaling ansatz,
\begin{eqnarray}
\varepsilon_0(\chi)-\varepsilon_0^*=C \chi^{-\kappa(1+z)},
\label{eq:eng_scale}
\end{eqnarray}
where $\varepsilon_0^*$ is the value of the energy per site in the limit $\chi\to\infty$,
$z$ is the dynamical critical exponent, and $C$ is a constant.
We investigate this behavior as a function of $\chi$ with $\varepsilon_0$ obtained by Eq.~(\ref{eq:varepsilon_0}).

\begin{figure}[t] 
\includegraphics[width=8.6cm, height=6.5cm]{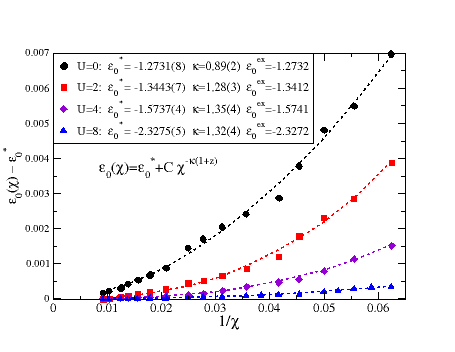}
\caption{The scaling behavior of the energy per site $\varepsilon_0(\chi)$ at $\mu=0$ for various $U$ 
as the matrix size $\chi$ changes. The consistency between $\varepsilon_0^*$, determined numerically by extrapolation of the fitting curves to the limit $\chi\to\infty$, and $\varepsilon_0^{ex}$,  calculated analytically based on the Bethe ansatz solution, confirms the validity of the scaling ansatz. We find that the value of $\kappa$ numerically determined
are well consistent with Eq.~(\ref{eq:kappa}) both for $c=2$ $(U=0)$ and for $c=1$ $(U>0)$.}
\label{fig:f1_fEm0}
\end{figure}
Figure~\ref{fig:f1_fEm0} shows $\varepsilon_0(\chi)$ at $\mu=0$ for various $U$. 
We determine $\varepsilon_0^*$ by extrapolating the fitting curves to the limit $\chi\to\infty$, and compare these values with
the Bethe ansatz solution\cite{Korepin05Book}
\begin{eqnarray}
\varepsilon_0^{ex}=-{U\over 4}-4\int_0^\infty \frac{d\omega}{\omega}\frac{J_0(\omega)J_1(\omega)}
{1+\exp(U\omega/2)}.
\label{eq:e1}
\end{eqnarray}
As presented in Fig.~\ref{fig:f1_fEm0}, $\varepsilon_0^*$ and $\varepsilon_0^{ex}$ are well consistent, confirming the validity of the scaling ansatz in Eq.~(\ref{eq:eng_scale}) as well as the method to find $\varepsilon_0(\chi)$, proposed in Eq.~(\ref{eq:varepsilon_0}). 
We also determine the exponent $\kappa$ from the fitting curves by adopting $z=1$ for the case\cite{Fisher89} with
the particle-hole symmetry at $\mu=0$.
The results are consistent with Eq.~(\ref{eq:kappa}) both for $c=2$ $(U=0)$ and for $c=1$ $(U>0)$ as shown in the figure.

Entanglement entropy has been recognized as the single most important quantity to capture
the entanglement properties of quantum systems.
It has been discussed that the half-chain EE of a one-dimensional critical system shows scaling behavior,
\cite{Calabrese04}
$S_h \sim (c/6) \log_2 \xi$, with a correlation length $\xi$.
In terms of the quantum correlation length $\xi_\chi$, we expect
\begin{eqnarray}
S_h=\frac{c\kappa}{6}\log_2 \chi + s_1,
\label{eq:EE}
\end{eqnarray}
where $s_1$ is a non-universal constant.
This prediction has been confirmed in the various one-dimensional models.\cite{Pollmann09,Tagliacozzo08,Pino12,Pirvu12,Wang15}

\begin{figure}[b] 
\includegraphics[width=8.6cm, height=6.5cm]{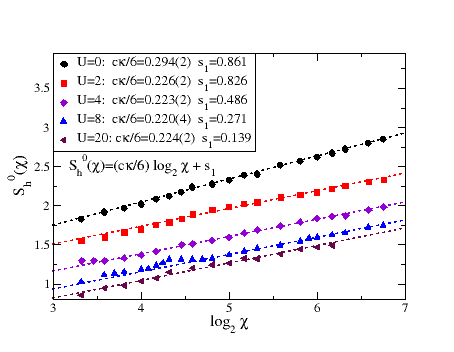}
\caption{The scaling behavior of the half-chain entanglement entropy $S_h^0(\chi)$ at $\mu=0$ 
as a function of $\chi$. The slope of the logarithmic dependence on $\chi$ determines $c\kappa/6$.
The resulting values of $\kappa$ quite strongly confirm Eq.~(\ref{eq:kappa}) again both for $c=1$ and for $c=2$.}
\label{fig:f2_fShm0}
\end{figure}
Thus the effects caused by a limited amount of entanglement counted in the MPS can be more explicitly investigated by measuring the half-chain EE
\begin{eqnarray}
S_h^0(\chi)= -\sum_{a=1}^\chi w_a \log_2 w_a,
\label{eq:fShm0}
\end{eqnarray}
where $w_a=\Lambda_a^2/\{\sum_{b=1}^\chi\Lambda_b^2\}$ are the eigenvalues of the reduced density matrix
of the half chain.
The eigenvalues are normalized by the condition $\sum_{a=1}^\chi w_a =1$ and usually ordered in the way
$w_1 \ge w_2 \ge \cdots$ so that $w_1$ is the largest eigenvalue.

The results are shown in Fig.~\ref{fig:f2_fShm0}.
The slope of $S_h^0(\chi)$ as a function of $\log_2\chi$ are determined by fitting data to Eq.~(\ref{eq:EE})
and yields the values for $c\kappa/6$.
Again the results shown in the figure are well consistent with Eq.~(\ref{eq:kappa}) both for $c=1$ ($c\kappa/6$=0220-0.226)
and for $c=2$ ($c\kappa/6$=0294) with error ranges as shown in the figure.

\subsection{Entanglement Spectrum}
Even though the EE clearly shows scaling behavior,
more properties beyond this single number can be revealed by the ES,
which, indeed, provides the universal properties of the entanglement entropy,
based on the scaling properties of the moments $R_n\equiv\text{Tr}\rho_A^n$.
These properties can be written as
\begin{eqnarray}
R_n=c_n L_{\text{eff}}^{-(c/12)(n-1/n)}, 
\label{eq:Rn}
\end{eqnarray}
according to the CFT calculations for $\rho_A$  of one-dimensional half-chain systems,\cite{Calabrese04}
where $c_n$ are non-universal constants, with $c_1=1$ by the normalization condition,
and $L_{\text{eff}}$ is a relevant length scale.

\begin{figure} 
\includegraphics[width=8.6cm, height=6.5cm]{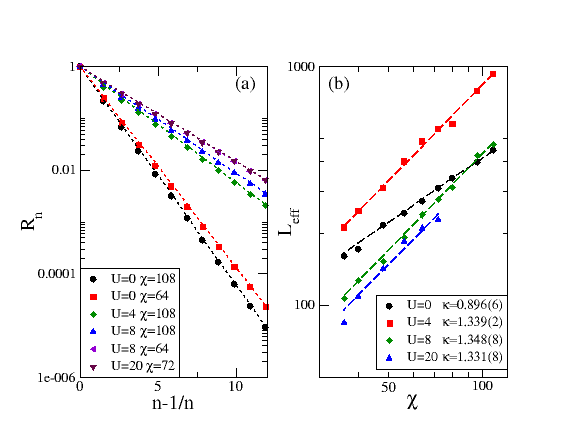}
\caption{(a) The $n$-th moments of the reduced density matrix for different $U$ and $\chi$ at half-filling.
The dotted lines represent $R_n=e^{-b(n-1/n)}$, where $b=-\ln w_1$ with $w_1$, the largest eigenvalue obtained numerically.
(b) The scaling behavior of the effective length, $L_{\text{eff}}\sim \chi^\kappa$, gives the critical exponents $\kappa$,
where $L_{\text{eff}}$ is obtained by fitting $R_n$ to the equation $R_n\sim L_{\text{eff}}^{-(c/12)(n-1/n)}$.}
\label{fig:f3_fRn}
\end{figure}
Figure~\ref{fig:f3_fRn}a represents $R_n$ for $c=2$ and $c=1$ in the one-dimensional Hubbard model.
It shows that  $\ln R_n$, nearly proportional to $n-1/n$ for whole rangle of $n$,
are very close to the dotted lines denoting $R_n=e^{-b(n-1/n)}$ with $b=-\ln w_1$.
The consistency between the lines and the numerical data supports the validity of the scaling properties of $R_n$ given in Eq.~(\ref{eq:Rn}) and the assumption $c_n \approx 1$, consistent with other works\cite{Pollmann10} in spin systems.
Note that, however,
different behavior of $c_n$ has been found\cite{Calabrese10} in $XXZ$ model, where $c_n$ decays exponentially with $n$.
In Fig.~\ref{fig:f3_fRn}b, $L_{\text{eff}}$ are determined by fitting data to Eq.~(\ref{eq:Rn}) for given $\chi$  and $U$.
Subsequently, by using $L_{\text{eff}}=\xi_0\chi^\kappa$ for a finite $\chi$ with a non-universal constant $\xi_0$,
we obtain $\kappa=0.896$ and $1.331-1.348$, with error ranges shown in the figure
for $c=2\ (\ U=0)$ and $c=1\ (\ U>0)$, respectively,
consistent again with Eq.~(\ref{eq:kappa}).

One way to quantify the ES is to represent it in terms of the eigenvalue distribution $P(w)=\sum_a \delta(w-w_a)$.
It has been claimed that $P(w)$ for one-dimensional systems in the critical regime provide an approximate distribution\cite{Calabrese08} determined by a single parameter.
The distribution can be derived by using Eq.~(\ref{eq:Rn})
with the assumption $c_n=1$, which is good in  this case as discussed above.
Since $\sum_{a=1}^\infty w_a=\int dw\ wP(w)=1$, we treat $wP(w)$ as a normalized density function.
Then, using the Stieltjes transform of $wP(w)$, 
we have\cite{Calabrese08}
\begin{eqnarray}
P(w)=\delta(w_1-w)+{2b\theta(w_1-w)}\frac{I_1(\xi_w)}{w\xi_w} 
\label{eq:Pw}
\end{eqnarray}
with $\xi_w\equiv2\sqrt{b\ln(w_1/w)}$, where $I_\alpha$ are the modified Bessel functions.
This equation, as expected, reproduces
$\int dw\ w^nP(w)=e^{-b(n-1/n)}$ 
for $n\ge1$.

One of the key elements that describe the properties of the ES is the mean number of eigenvalues larger
than a given $w$, defined by
\begin{eqnarray}
n(w)\equiv\int_w^{w_1} du\ P(u).
\label{eq:n0}
\end{eqnarray}
Inserting Eq.~(\ref{eq:Pw}) into Eq.~(\ref{eq:n0}), then, we have
\begin{eqnarray}
n(w)=I_0(\xi_w)
\label{eq:nw}
\end{eqnarray}
\begin{figure}[t] 
\includegraphics[width=8.6cm, height=6.5cm]{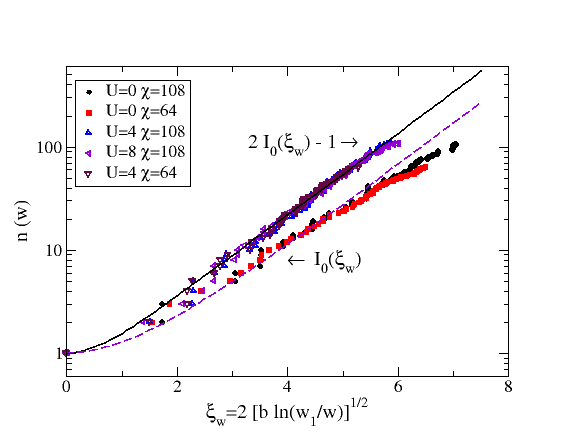}
\caption{The mean number of eigenvalues larger than a given $w$, $n(w)$, obtained for different $U$ and $\chi$ at half-filling.
For $c=2$ at $U=0$,  $n(w)$ shows the behavior $n(w)=I_0(\xi_w)$ with $\xi_w\equiv 2[b\ln(w_1/w)]^{1/2}$, as expected by the CFT calculations, while for $c=1$ in the region $U>0$, $n(w)$ behaves as $n(w)= 2I_0(\xi_w)-1$,
showing that a global double-degeneracy in the ES asymptotically occurs.}
\label{fig:f4_fn0}
\end{figure}

The prediction for $n(w)$ has been checked numerically in $XXZ$ spin models,\cite{Calabrese08,Pollmann10,Alba12,Laflorencie14},
confirming that Eq.~(\ref{eq:nw}) works well for the isotropic $XX$ point, but holds a sizable deviation for some anisotropic points. 
Here, we check the prediction in our model for both $c=1$ and $c=2$.
Figure~\ref{fig:f4_fn0} shows $n(w)$ obtained by counting the number of the eigenvalues $w_i$ in the MPS with finite matrices of size $\chi$. The parameter $b$ is chosen from the largest eigenvalue in the ES numerically obtained.
For $U=0$, $n(w)$ satisfies well the expectation in Eq.~(\ref{eq:nw}) in the wide range of $\xi_w$.
We believe that a weak deviation in the area of large $\xi_w$ is caused by the limited size of $\chi$. 
For $U>0$, however, we have $n(w) \sim 2 I_0(\xi_w)-1$.
This is a manifestation of the global degeneracy in the ES, discussed recently\cite{Alba18},
which leads to a modified formula $n(w)\sim g I_0 (\xi_w)$ for the ES with a global degeneracy $g$.

\begin{figure}[b] 
\includegraphics[width=8.6cm, height=6.5cm]{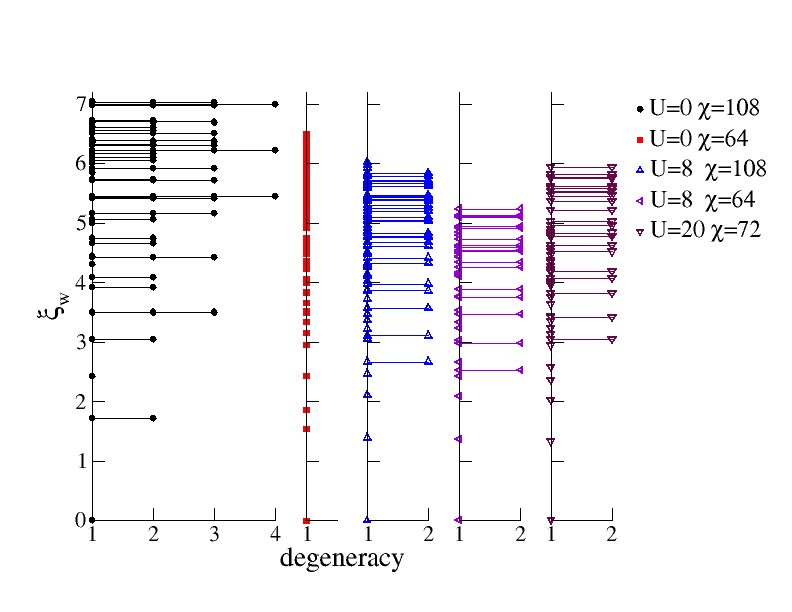}
\caption{The eigenvalue distribution of the reduced density matrices of the MPS ground states.
For $c=2$ at $U=0$,  no global degeneracy happens in spite of some accidental degeneracies.
However, for $c=1$ in the region $U>0$, asymptotically global double degeneracy appears, more distinctly
at larger $U$.
Note that no degeneracy appears in strong entanglement region (for small $\xi_w$), which is a hallmark of the Mott phase.}
\label{fig:f5_fES}
\end{figure}
In order to check the global degeneracy, we explicitly display the ES in Fig.~\ref{fig:f5_fES}.
We find that for large eigenvalues (i.e. small $\xi_w$), including the largest eigenvalue ($\xi_w=0$), no degeneracy
appears, which is a hallmark of the Mott phase with a charge excitation gap.
Namely, the resulting ground state has the antiferromagnetic ordering and the eigenstate with the largest eigenvalue of the reduced
density matrix state for the half-chain system is not degenerate. 
Note that at $U=0$ the ground state with a finite $\chi$ favors the Mott phase rather than a metallic phase. 
In this case, furthermore, no global degeneracy happens ignoring some accidental degeneracies.
However, for $U >0$, asymptotically a global double-degeneracy appears in the region of weak entanglement 
(i.e. for rather large $\xi_\chi$).
This tendency becomes more eminent for larger $U$.
In fact, it is not easy to decide whether the global double-degeneracy appears for $U > 0$
just by looking at the distribution of the eigenvalues since single and double degeneracies are mixed.
Therefore, $n(w)$, which shows quite distinct asymptotic behavior, is indeed a very convenient tool to figure out the global degeneracy.
Indeed, the ES with the global double-degeneracy in weak entanglement region for $U > 0$ 
can be regarded as a fingerprint of the state antiferromagnetically ordered, 
in which low-energy spin excitations are gapless while high-energy charge excitations are gapped.

\begin{figure}[b] 
\includegraphics[width=8.6cm, height=6.5cm]{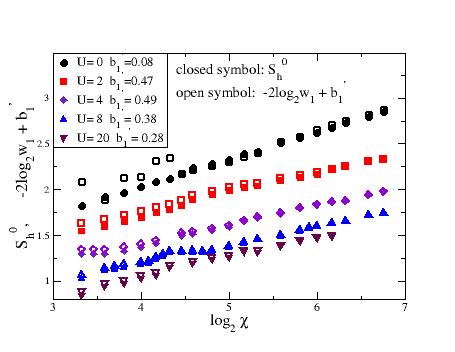}
\caption{The comparison between the half-chain entanglement entropy $S_h^0$ and the single-copy entanglement $-\log_2 w_1$ supports the relation $S_h^0 = -2\log_2 w_1+ b_1$, with a non-universal parameter $b_1$ which is determined numerically.
}
\label{fig:f6_fShbm0}
\end{figure}
From the R\'enyi entropy $S_n\equiv \frac{1}{1-n}\log_2 R_n$ , we have
$S_h^0\sim -2\log_2 w_1$ with $R_n=e^{-b(n-1/n)}$.
This means that $-\log_2 w_1$, the so-called single-copy entanglement\cite{Eisert05,Peschel05,Orus06}
which can be distilled from a single specimen of the quantum system,  
is half of the half-chain EE which can be obtained from many identically prepared systems.
Therefore, it is interesting to check this relation in our MPS of the one-dimension Hubbard model.
Fig.~\ref{fig:f6_fShbm0} shows this behavior supporting the relation 
$S_h^0 = -2\log_2 w_1+ b_1$ with a non-universal parameter $b_1$ which is determined numerically for a good agreement
between these two quantities.

\section{First-order nature of the metal-insulator transition for $\mu_c <0$}
\begin{figure}[t] 
\includegraphics[width=8.6cm, height=6.5cm]{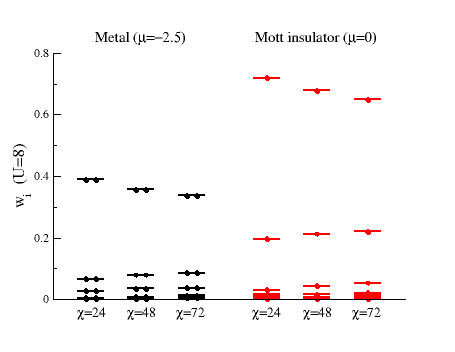}
\caption{A few largest eigenvalues of the reduced density matrix in the metallic and the Mott-insulating phases.
The metallic phase clearly shows double-degeneracy in the whole ES denoted by two dots in the figure,
distinctly different from the non-degenerate Mott-insulating phase denoted by single dots.}
\label{fig:f7_fLs}
\end{figure}

We investigate the nature of the transition between the metallic and the Mott-insulating phases in the MPS with a finite $\chi$
by tuning the chemical potential $\mu$ in the region below half-filling ($\mu < 0$) .
The two phases can be identified by the degeneracy in the ES as shown in Fig.~\ref{fig:f7_fLs}:
the metallic phase clearly shows doubly degenerate ES for the whole range of the eigenvalues
while the Mott-insulating phase shows non-degenerate ES for large eigenvalues.
The double-degeneracy in the metallic phase reflects the $Z_2$ symmetry for the parity of the fermionic particle number.
In the insulating phase, the Schmidt gap\cite{Chiara12}, defined as $w_1-w_2$, is finite and is expected to be vanishing
only in the limit $\chi\to\infty$.
This causes a first-order transition between these two phases for a finite $\chi$
because the ES cannot be smoothly connected across the transition point.  

\begin{figure}[t] 
\includegraphics[width=8.6cm, height=6.5cm]{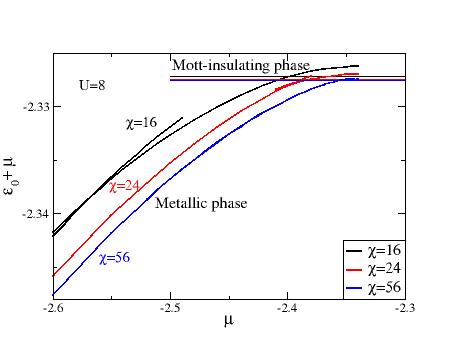}
\caption{The energy curves for the metallic and the Mott-insulating phases with a given $\chi$ as a function of the chemical
potential $\mu$. In some metallic regions, there are multiple curves. The curves cross each other,
showing the nature of first-order transitions.}
\label{fig:f8_fe0}
\end{figure}
The transition point is determined by comparing the energies of the MPS for two phases as plotted in Fig.~\ref{fig:f8_fe0}.
As a function of the chemical potential, $\varepsilon_0+\mu$ is constant in the Mott-insulating phase.
In the metallic phase, we perform the TEBD process by using the final MPS for a given $\mu$ as the initial trial wavefunction
for the next stage with $\mu$ slightly changed.
In this way, we gradually change $\mu$ until the value of the energy deviates from a smooth curve of $\varepsilon_0(\mu)$.
The energy curves for the metallic and the Mott-insulating phases cross at a first-order transition point when $\chi$ is rather small.
\cite{Sandvik07}
Furthermore, in some regions of the metallic phase, there are multiple curves for $\varepsilon_0(\mu)$ 
crossing each other as shown in the figure,
which is a typical phenomenon in the variational solutions, like the MPS solutions, with a finite number of parameters.
This means that there are first-order transitions from a metallic phase to a different metallic phase in the MPS with a finite $\chi$.

\begin{figure}[b] 
\includegraphics[width=8.6cm, height=6.5cm]{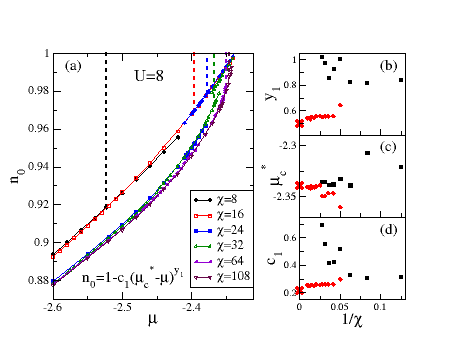}
\caption{(a) The particle density, $n_0$, as a function of $\mu$ for the MPS with a given $\chi$
shows different asymptotic behavior depending on $\chi$ as approaching to the insulating phase.
The dotted lines represent the metal-to-insulator transition points.
(b) The critical exponents governing the asymptotic behavior of $n_0$.
The values are divided into two groups showing mean-field nature (red diamond) approaching to $y_1=1$ and strongly correlated nature (black square) approaching to $y_1=1/2$ as $\chi$ increases. 
(c) The transition points extracted from the curves.
(d) Amplitudes in the asymptotic behavior of $n_0$.}
\label{fig:f9_fn}
\end{figure}
Now we investigate the transition from the metal to the Mott-insulator by measuring the particle density, $n_0$,
as a function of $\mu$.
Grand canonical calculations in an infinite system allow us to change $n_0$ continuously as $\mu$ changes.
In the Mott-insulating phase, the density keeps constant to be $n_0=1$.
In the metallic phase, $n_0$ changes as a function of $\mu$ and becomes closer to $n_0=1$ as approaching to the insulating phase.
Figure~\ref{fig:f9_fn}a shows this behavior of $n_0$ for $U=8$.
Obviously, the curves for $n_0(\mu)$ exhibit different asymptotic behaviors depending on $\chi$, even though
the first-order transitions occur at crossing points of the energies.
The asymptotic properties of $n_0$ can be investigated by expressing it in the form
\begin{eqnarray}
n_0=1- c_1 (\mu_c^* - \mu)^{y_1},
\end{eqnarray}
where $y_1$ is an exponent characterizing the power-law behavior,
$\mu_c^*$ is an effective transition point of the curves, and $c_1$ is an amplitude.
The values of these three parameters determined by fitting for given $\chi$ are shown in Fig.\ref{fig:f9_fn}b--d.
Note that the exact values\cite{Korepin04} based on the Bethe ansatz solutions are $y_1^{ex}=1/2$, $\mu_c^{ex}=-2.340$,
and $c_1^{ex}=0.215$.

Fig.\ref{fig:f9_fn}b--d show that there are two groups of curves denoted by squares (black) and diamonds (red), whose $y_1$
approaches to 1 (the mean-field value) and to 1/2 (the exact value), respectively, as $\chi$ increases.
This implies that when the correlation length is larger than $\xi_\chi$, we always have mean-field
solutions (black squares) with effective transition points $|\mu_c^*| < |\mu_c^{ex}|$, 
consistent with the expectation that in the limit $\chi\to 0$ the mean-field solution have $\mu_c^*\to 0$.
On the other hand, the other group of curves (red diamonds), say, the strongly-correlated solutions, have $|\mu_c^*| > |\mu_c^{ex}|$.
The parameters $y_1$, $\mu_c^*$, and $c_1$ for these curves converge to the exact values as $\chi$ increases.

\begin{figure}[b] 
\includegraphics[width=8.6cm, height=6.5cm]{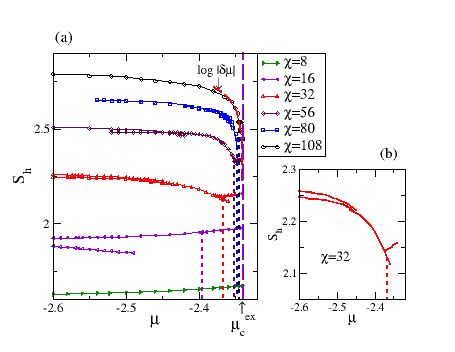}
\caption{(a) The half-chain EE as a function of the chemical potential $\mu$ in the vicinity of the transition point $\mu_c^{\text{ex}}=-2.340$ for various $\chi$. The dotted lines mark the first-order transition points for given $\chi$.
(b) A enlarged figure of the curves for $\chi=32$ provides a better view for the existence of multiple MPS solutions
showing the strongly correlated as well as the mean-field nature. }
\label{fig:f10_fSh}
\end{figure} 
The half-chain EE in the metallic phase, $S_h$, also shows the mean-field nature close to the critical point.
Figure~\ref{fig:f10_fSh}a represents the half-chain EE for various $\chi$.
The asymptotic behavior of the strongly-correlated solutions show $S_h\sim \log_2 |\mu_c^*-\mu|$,
whereas the mean-field solutions appear deviating from the logarithmic behavior close to the critical point.
This can be more explicitly observed in Fig.~\ref{fig:f10_fSh}b of the enlarged curves for $\chi=32$.
There are multiple curves of the strongly-correlated solutions, roughly logarithmically changing, in the metallic regions
and first-order transitions between them.
Near the critical point, however, curves appear deviated from the strongly-correlated solutions,
which is smoothly increasing as approaching to the insulating phase.
Those curves, the mean-field solutions for $\chi=32$, have larger energy than those of the strongly correlated solutions
at the transition point. 
For smaller $\chi$'s, however, the mean-field solutions appear in wider range of $\mu$ as the ground states.
In this case, the first-order transitions occur between the mean-field metallic state to the Mott-insulating state.

\section{Summary}
We study the effects caused by limited amount of entanglement on the ground states of the one-dimensional Hubbard model
by adopting the MPS representations with finite size matrices.
The two-site TEBD method is used to optimize this variational MPS.
As a function of the matrix size $\chi$, we find
that the finite-entanglement effects can be characterized by an effective correlation length $\xi_\chi \sim \chi^\kappa$.
The finite-$\chi$ scaling behavior of the energy and the half-chain entanglement entropy at half-filling provides the values of $\kappa$ consistent with the theoretical prediction based on the CFT for both $c=1 \ (U >0)$ and $c=2 \ (U=0)$.
The entanglement spectrum also shows a distribution consistent with the CFT prediction for $U=0$ while the case for $U>0$
shows an occurrence of a global double-degeneracy in the Mott phase.
In the MPS with finite size of matrices, the metal-to-insulator transitions always occur through a first-order transition. 
Furthermore, multiple variational solutions exist in the metallic phase, including ones showing the mean-field nature
close to the critical point.

\begin{acknowledgments}
The author greatly appreciates helpful comments from Pasquale Calabrese, Ian McCulloch and Zhiyuan Xie,
and useful discussions with Myung-Hoon Chung and Ji-Woo Lee.
This work was supported by the Basic Science Research
Program through the National Research Foundation of Korea
funded by the Ministry of Education, Science and Technology (Grant No. NRF-2016R1D1A1B03935815).
\end{acknowledgments}

\end{document}